\documentclass[10pt]{IEEEtran}

\usepackage{shortcuts_tmr}

\usetikzlibrary{external}
\tikzsetexternalprefix{figs/tikz/precompiled/}

\def\cH{{\cal H}}
\def\A{{\mathbf A}}

\def\I{{\mathbf I}}

\def\Cy{{\mathbf C}_y}
\def\Cym{\widehat{\mathbf C}_y^m}

\def\wl{{\mathbf w}^{\paren{\ell}}}
\def\yl{{\mathbf y}^{\paren{\ell}}}

\def\u{{\mathbf u}}
\def\uhat{\widehat{\u}}
\def\utilde{\widetilde{\u}}
\def\v{{\mathbf v}}
\def\vhat{\widehat{\v}}

\def\e{{\mathbf e}}
\def\x{{\mathbf x}}
\def\y{{\mathbf y}}
\def\beps{\boldsymbol{\epsilon}}
\def\bbone{{\mathbf 1}}

\def\G{{\cal G}}
\def\Verts{{\cal V}}
\def\Edges{{\cal E}}

\def\relates{\sqsupset}
\def\relatesstar{{\sqsupset^\ast}}

\title{Blind Inference of Eigenvector Centrality Rankings}
\author{
  T.~Mitchell~Roddenberry and Santiago~Segarra\\
  \thanks{
    T.M. Roddenberry and S. Segarra are with the Dept. of ECE, Rice University.
    Work in this paper is supported by the USA NSF CCF-2008555 and the Ken Kennedy 2019/20 AMD Graduate Fellowship.
    Emails: mitch@rice.edu, segarra@rice.edu.
    Preliminary results appeared in a conference publication~\cite{roddenberry2020ranking}.
  }
}

\begin{document}

\maketitle

\begin{abstract}
  We consider the problem of estimating a network's eigenvector centrality only from data on the nodes, with no information about network topology.
Leveraging the versatility of graph filters to model network processes, data supported on the nodes is modeled as a graph signal obtained via the output of a graph filter applied to white noise.
We seek to simplify the downstream task of centrality ranking by bypassing network topology inference methods and, instead, inferring the centrality structure of the graph directly from the graph signals.
To this end, we propose two simple algorithms for ranking a set of nodes connected by an unobserved set of edges.
We derive asymptotic and non-asymptotic guarantees for these algorithms, revealing key features that determine the complexity of the task at hand.
Finally, we illustrate the behavior of the proposed algorithms on synthetic and real-world datasets.
\end{abstract}

\section{Introduction}\label{sec:intro}

As relational, non-Euclidean data has become increasingly prominent, so has the need for algorithms to make sense of arbitrarily structured datasets.
The representation of data as graphs, or networks, is a popular approach~\cite{strogatz2001,newman2010,jackson2010}, allowing one to uncover community structure~\cite{newman2006modularity}, common connection patterns~\cite{milo2002networkmotifs}, and node importance~\cite{pagerank1999}.

The topology of a network commonly reflects mutual influence between the individual nodes.
Hence, to concisely understand graph properties, ranging from behavior of dynamics on the graph to vulnerability to external attacks, practitioners often employ \emph{centrality measures} to identify the most influential or important nodes in the network topology.
Given that this abstract notion of importance can be formalized in different ways, a wide range of centrality measures co-exist in the literature including closeness~\cite{beauchamp1965improved}, betweenness~\cite{freeman1977set,segarra2016stability}, and eigenvector~\cite{bonacich1972factoring} centralities.
Due to its widespread use~\cite{friedkin1991theoretical,fletcher2018structure,landau1914spielturnieren}, in this work we are primarily concerned with the eigenvector centrality, where the importance of each node is dependent on the importance of its neighbors.

The eigenvector centrality value for each node in the network is given by the corresponding entry in the leading eigenvector of the graph's (potentially weighted) adjacency matrix, defined later in~\cref{sec:prelim:graphs}.
Thus, to compute these values, one requires full knowledge of the network topology.
In resource-constrained settings, this information can be expensive or even impossible to obtain.
Moreover, some scenarios may not have a clear network to analyze, requiring one to be constructed using network topology inference techniques.
These methods leverage data (or graph signals) supported on the nodes of the graph, and infer the set of edges by assuming the graph signals are regularized by the underlying graph structure.

As a motivating example, consider a social network where the set of edges (friendships) is not known precisely, but one has snapshot measurements of opinion dynamics among all of the individuals.
One then seeks to infer who the most central individual in the network is.
In the network topology inference framework, one would use this collection of opinions to infer a graph structure, and then compute the eigenvector centrality of the constructed network.
This approach requires the costly -- both in the data and computational sense -- construction of an intermediate network, although the main interest is just in the resulting centrality.
This leads to the core question of this work:
\emph{Under what conditions can the nodes of a graph with hidden edges be ranked according to their eigenvector centrality directly from data supported on them?}
Working in the framework of graph signal processing, we model this data as a set of graph signals obtained from an unknown network process, which we characterize as a graph filter applied to white noise.
We then reveal how the difficulty of this problem is determined by the distribution of centrality over the nodes and the spectral properties of the graph and graph filter.

\subsection{Related literature}\label{sec:intro:lit}

In the canonical setting of the well-studied problem of \emph{network topology inference}, data on the nodes is used to infer the complete set of edges.
Particularly, in the context of inferring graphical models, each node represents a random variable and edges encode the conditional dependence structure between random variables~\cite{friedman2007lasso,lake2010discovering,meinshausen2006high}.
The inference of this structure from the covariance of the observed random variables is often complemented with sparsity assumptions, yielding a graph-structured model.

Closer to our current work, graph signal processing provides a different view on network topology inference, where the nodal data is assumed to be the output of a latent process on the hidden graph~\cite{dong2016laplacian,kalofolias2016smooth,segarra2017topo,mateos2019connecting}.
In this way, topology inference is cast as an inverse problem where we seek to recover the network structure from the observed output.
To overcome the inherent ill-definition of this inverse problem, existing approaches make different assumptions on the process that generates the graph signals, \eg kernel models~\cite{shen2017kernel}, signal smoothing~\cite{dong2016laplacian}, or consensus dynamics~\cite{zhu2019network}.

Motivated by the elevated sampling and computational demands of network topology inference, recent works have considered the problem of \emph{blind} community inference, i.e., estimating the communities in a graph directly from the observation of graph signals.
The observation of a simple finite-length diffusion process with white noise input on a planted partition graph was studied in~\cite{schaub2018dynamical}, where the authors characterized the relationship between the diffusion time and the difficulty of identifying the latent communities.
Additional models for the observed signals have been considered including a low-rank generation process~\cite{wai2018lowrank} and a function of a latent time series~\cite{hoffmann2018unobserved}.
Scenarios where the underlying graphs are time-varying but preserve a common community structure have also been studied~\cite{roddenberry2020community}.

In the direction of centrality estimation without complete edge data, \cite{ruggeri2019centrality} infers eigenvector centralities on networks where some edges are unobserved, but do not consider nodal data.
\cite{shao2017centrality} infers a temporal centrality based on nodal data, but does not consider eigenvector centralities or account for arbitrarily filtered graph signals.
Most closely related to this work, \cite{he2020centrality} considers the estimation of eigenvector centralities when the excitation signal to the network process is colored, \ie not white noise.
However, they are concerned with estimating eigenvector centralities \emph{precisely}, which they characterize by providing norm-bounds on the eigenvector estimation error.
In contrast, this work is concerned with \emph{ranking} nodes, which we characterize in terms of sampling requirements for particular pairs of nodes~\cf{\cref{thm:simple-nonasymptotic}}.

\subsection{Contributions and outline}\label{sec:intro:cont}

\noindent The contributions of this paper are threefold:
\begin{enumerate}
\item We provide two simple algorithms for estimating the ranking of nodes according to their eigenvector centrality, based exclusively on graph signals.
\item We guarantee the correctness of these algorithms in the asymptotic regime and derive requirements for desired performance of each algorithm in the non-asymptotic regime, thus revealing key features that determine the complexity of the task at hand.
\item We illustrate the performance of the algorithms and validate our theoretical results with experiments on synthetic and real-world data.
\end{enumerate}

In~\cref{sec:prelim}, we gather notation, definitions, and introduce graph signal processing concepts.
In~\cref{sec:problem}, we define our system model along with relevant assumptions and we provide a formal definition of the ranking problem to be studied.
A simple algorithm is put forth in~\cref{sec:warmup} along with theoretical results characterizing its asymptotic and non-asymptotic performance.
\cref{sec:ranking-prior} presents and analyzes a refined version of the previous algorithm that takes into account the effect of sampling noise in the estimate of the centrality ranking.
We demonstrate our results on synthetic and real data in~\cref{sec:experiments}, making special emphasis on the effects of sample size and graph structure on ranking performance.
Finally, a concluding discussion and directions for future research close the paper in~\cref{sec:discussion}.

\section{Preliminaries}\label{sec:prelim}
  
\subsection{General notation}\label{sec:prelim:notation}

The entries of a matrix $\mathbf X$ and vector $\x$ are referred to by $X_{ij}$ and $x_{ij}$, respectively.
The collection of integers $\brack{1,\ldots,n}$ is denoted by $\sbrack{n}$.
The notation $\norm{\cdot}_2$ indicates the standard $\ell_2\text{-norm}$ for vectors, and the $\ell_2\text{-operator norm}$ for matrices.
$\I$ indicates the identity matrix, and $\bbone$ indicates the all-ones vector.
For a vector $\x\in\Real^n$, $\x^\top\in\Real^{1\times n}$ indicates the transpose of $\x$.
The notation $\E\sbrack{\cdot}$ indicates the expected value of a random variable, and $\Cov\paren{\cdot}$ denotes the covariance matrix of a random vector.
For two vectors $\x,\y$, $\cos\theta\paren{\x,\y}={\abrack{\x,\y}}/\paren{\norm{\x}_2\norm{\y}_2}$, and $\sin\theta\paren{\x,\y}=\sqrt{1-\cos^2\theta\paren{\x,\y}}$.
The notations ${\cal O}\paren{\cdot}, \Omega\paren{\cdot}, \Theta\paren{\cdot}$ denote the established asymptotic equivalence classes for functions~\cite[Chapter 3]{cormen2009algorithms}.
Finally, the binary relation of an ordering is indicated by $\relates$.

\subsection{Graphs and eigenvector centrality}\label{sec:prelim:graphs}

An undirected \emph{graph} $\G$ consists of a set $\Verts$ of $n\deq\abs{\Verts}$ nodes, and a set $\Edges\subseteq\Verts\times\Verts$ of edges, corresponding to unordered pairs of nodes.
Such a structure is compactly represented by an \emph{adjacency matrix} $\A$, generated by an arbitrary indexing of the nodes with the integers $1,\ldots,n$, and then assigning
\begin{equation}\label{eq:adjacency-matrix}
  A_{ij} =
  \begin{cases}
    1 & \Paren{i,j}\in\Edges, \\
    0 & \text{otherwise}.
  \end{cases}
\end{equation}

The \emph{eigenvector centrality}~\cite{beauchamp1965improved} of a node $i$ in a connected graph $\G$ is given by the $i^{\rm th}$ entry of the leading eigenvector $\v_1$ of $\A$.
The Perron-Frobenius Theorem guarantees that every element of $\v_1$ has the same sign, since every element of $\A$ is non-negative.
To handle the sign-ambiguity present in eigenvectors, we denote the ``positive-signed'' orientation of $\v_1$ with $\u$.

In many applications, we are not interested in particular values of $\u$, but rather the \emph{relative} centralities of nodes.
More precisely, we concern ourselves with the \emph{centrality ranking}, characterized by the endowment of the set of nodes $\Verts$ with a \emph{weak ordering} $\relatesstar$ obeying the following:
\begin{equation}\label{eq:centrality-ranking}
  \text{for all } i,j\in\Verts\colon u_i< u_j\iff i \, \relatesstar j.
\end{equation}
We specify this as a weak ordering since two different nodes may have the same eigenvector centrality, \eg two isomorphic nodes in $\G$.
{In such cases, the two nodes are `tied,' so neither $i\,\relatesstar j$ nor $j\,\relatesstar i$.}

\subsection{Graph signals and graph filters}\label{sec:prelim:gsp}

\emph{Graph signals}, analogously to discrete time signals, are real-valued functions on the nodes, \ie $x\colon\Verts\to\Real$.
For an indexing of $\Verts$ with $\sbrack{n}$, a graph signal $\x$ is represented as a vector in $\Real^n$, where $x_i=x\paren{i}$.
A graph filter $\cH$ is a linear map between graph signals representable as a polynomial of the adjacency matrix\footnote{In general, a graph filter can be written as a polynomial of any matrix that captures the graph structure, usually denominated by graph shift operator. In this paper, we focus on the particular case of the adjacency matrix.}
\begin{equation}\label{eq:graph-filter}
  \cH\!\Paren{\A} = \sum_{k=0}^T\gamma_k\A^k \deq \sum_{l=0}^n\cH\!\Paren{\lambda_l}\v_l\v_l^\top,
\end{equation}
where $\gamma_k$ are real-valued coefficients, $\cH\paren{\lambda}$ is the extension of the polynomial $\cH$ to scalar-valued arguments, and $\paren{\lambda_l,\v_l}$ denote the eigenpairs of $\A$. The filter definition in~\eqref{eq:graph-filter} can model a wide gamut of phenomena on graphs, including diffusion processes~\cite{Masuda2017}, consensus dynamics~\cite{Olfati-Saber2007}, and a variety of biological processes~\cite{huang2018fmri}.

\section{Problem Setting}\label{sec:problem}

\subsection{System model}\label{sec:problem:model}

Consider a set of $m$ graph signals obtained as the output of an unknown graph filter.
Formally, for each $\ell\in\sbrack{m}$, we observe
\begin{equation}\label{eq:system-model}
  \yl \deq \cH\!\Paren{\A}\wl,
\end{equation}
where $\cH$ is a polynomial of the graph adjacency matrix $\A$, and the set $\{\wl\}_{\ell=1}^m$ consists of \iid\ samples from a zero-mean distribution obeying $\Cov \paren{\wl} \deq \mathbb{E} [\wl \paren{\wl}^\top]=\I$.
We emphasize that $\A$ and $\cH$ are unknown in this scheme, and only $\{\yl\}_{\ell=1}^m$ are observed.
In the motivating example of a social network, one would observe snapshots of opinion dynamics over a population with \emph{no knowledge} of the underlying social network~$\A$ or the specific opinion formation process~$\cH$.
From these opinion snapshots, commensurate with the opinion dynamics of the network structure, one seeks to uncover the most influential person in the population.

Observe that by considering \eqref{eq:graph-filter}, the covariance matrix of the signals following~\eqref{eq:system-model} shares the set of eigenvectors with the adjacency matrix.
Specifically,
\begin{equation}\label{eq:shared-eigenvectors}
  \begin{split}
    \Cy &\deq \Cov\!\Paren{\yl} = \cH\!\Paren{\A}\Cov\!\Paren{\wl}\cH\!\Paren{\A} \\
    &= \Sbrack{\cH\!\Paren{\A}}^2 = \sum_{l=0}^n\Sbrack{\cH\!\Paren{\lambda_l}}^2\v_l\v_l^\top,
  \end{split}
\end{equation}
where we have used the facts that $\A$ is a symmetric matrix and that the covariance of $\wl$ is the identity matrix.
Hence, one can analyze the spectral structure of a graph strictly from the observation of such signals, without knowledge of the graph itself.
Throughout our analysis, we focus on the following class of graph filters:
\begin{assump}\label{ass:filter-invertability}
  The graph filter $\cH$ in~\eqref{eq:system-model} has non-negative coefficients $\gamma_k$~\cf{\eqref{eq:graph-filter}}.
\end{assump}
It follows from~\eqref{eq:shared-eigenvectors} that \Cref{ass:filter-invertability} guarantees that the leading eigenvector of $\Cy$ is unique and equal to $\u$ (up to sign). This assumption is reasonable, corresponding to the process described by $\cH$ \emph{constructively} aggregating information between nodes.
That is, the nodes in any filter $\cH$ obeying \Cref{ass:filter-invertability} have a \emph{positive influence} on each other.
Additionally, we make the following assumption on the distribution of the observed signals $\yl$.
\begin{assump}\label{ass:bounded-signals}
  There exists $r>0$ such that $\norm{\yl}_2\leq r$ almost surely.
\end{assump}
\Cref{ass:bounded-signals} bounds the magnitude of the observed graph signals $\yl$, generated according to~\eqref{eq:system-model}.
This assumption is not restrictive, as it can be satisfied by the setting where $\wl$ has bounded norm and $\cH\paren{\A}$ has a bounded spectral radius.
\begin{remark}[Relaxing Assumptions~\ref{ass:filter-invertability} and~\ref{ass:bounded-signals}]\label{remark:selection}
  Given a graph filter not abiding by \cref{ass:filter-invertability}, the leading eigenvector of $\Cy$ may be a poor estimate of the graph's eigenvector centrality $\u$.
  However, it is still the case that $\Cy$ and $\A$ share the same set of eigenvectors.
  Thus, one could leverage the fact that $\u$ has same-signed entries~\cf{\cref{sec:prelim:graphs}} to select the correct estimator of $\u$ from within the spectrum of $\Cy$, rather than simply choosing the leading eigenvector.
  Many of our presented results still hold in this setting, but we maintain~\cref{ass:filter-invertability} for simplicity.
  We refer to~\cite{roddenberry2020selection} for discussion on the problem of selecting the correct vector in this non-ideal setting.
  Regarding \cref{ass:bounded-signals}, although our theoretical results rely on it, they can be extended to the more general case where $\yl$ follows a subgaussian distribution mimicking the development in~\cite[Corollary 5.50]{vershynin2010introduction}.
\end{remark}

\subsection{Eigenvector centrality ranking}\label{sec:problem:ranking}

We seek to \emph{order} the nodes of a graph according to their eigenvector centrality, as formally stated next.  
\begin{problem}\label{prob:ranking}
  Given the observation of $m$ graph signals following the system model in~\eqref{eq:system-model}, rank the nodes of the graph according to the eigenvector centrality $\u$.
\end{problem}

By solving~\cref{prob:ranking}, we do not necessarily aim to infer the \emph{precise} eigenvector centrality of each node in the graph.
Rather, by using indirect information about the graph structure obtained via graph signals, we aim to discern the relative ranking of nodes in the eigenvector centrality.
Although the graph structure -- from which the eigenvector centrality is computed -- is not observed directly, the graph signals \emph{do} encode the influence structure of the network.
To illustrate this, consider a more favorable setting where both the inputs $\wl$ and outputs $\yl$ are observed.
Intuitively, one would expect the nodes with the highest eigenvector centrality to have the most influence in the output $\yl$.
Thus, by measuring the influence of a node's input value on the output $\yl$, the eigenvector centrality ranking should be discernible.

However, in the more challenging setting considered in \cref{prob:ranking}, the input signals $\wl$ are \emph{not observed}.
To circumvent this challenge, we may leverage the fact that the covariance matrix $\Cy$ shares the set of eigenvectors with $\A$, where both matrices have the same leading eigenvector $\u$ under~\cref{ass:filter-invertability}.
Hence, we may estimate $\u$ by taking the leading eigenvector of the sample covariance matrix of the graph signals $\{\yl\}_{\ell=1}^m$.

Given finitely many samples, the true covariance matrix $\Cy$ is not available.
So, we need to understand the effects of finite sampling on the eigenspaces of $\Cy$ and $\Cym\deq\frac{1}{m}\sum_{i=1}^m\yl[\yl]^\top$.
If $\Cym$ is a reasonably good estimate of $\Cy$, one expects the weak ordering $\relates$ induced by $\uhat$, the leading eigenvector of $\Cym$, to be close to the sought order $\relatesstar$, induced by the leading eigenvector of $\Cy$.
Indeed, if the difference between $\u$ and $\uhat$ is sufficiently small, the weak orderings $\relates$ and $\relatesstar$ will be equivalent.

On a finer scale, consider~\cref{prob:ranking} in a pairwise sense, where we are only concerned with the ordering of two nodes $i,j\in\Verts$.
If $\abs{u_i-u_j}$ is large, it is reasonable to expect that order to be conserved under the estimated eigenvector $\uhat$.
Inversely, if $\abs{u_i-u_j}$ is small, \cref{prob:ranking} is expected to be difficult for nodes $i,j$.

{Considering the estimated $\relates$ and true $\relatesstar$ orderings, we say that $\relates$ and $\relatesstar$ are \emph{concordant} with respect to a pair of nodes $i,j\in\Verts$ where $i\, \relatesstar j$  if $i\relates j$.
They are \emph{discordant} with respect to $i,j$ if $j\relates i$.
If $i,j$ are tied, \ie neither $i\,\relatesstar j$ nor $j\,\relatesstar i$, then $\relates$ must also tie $i,j$ to be concordant with respect to $i,j$, otherwise it is discordant.}

Motivated by this, we characterize the difficulty of~\cref{prob:ranking} in terms of the number of samples $m$ needed to correctly order two nodes relative to each other.

\section{Warm-up: Simple ranking}\label{sec:warmup}

\begin{algorithm}[tb]
  \caption{Blind centrality ranking}\label{alg:simple}
  \begin{algorithmic}[1]
    \STATE {\bf Input:} Graph signals $\brack{\yl}_{\ell=1}^m$
    \STATE Compute the sample covariance matrix
    \begin{equation}\label{eq:alg:simple:covariance}
      \Cym \deq \frac{1}{m}\sum_{i=1}^m \y^{\Paren{i}}\Sbrack{\y^{\Paren{i}}}^\top
    \end{equation}
    \STATE Compute the eigenvector decomposition of $\Cym$
    \begin{equation}\label{eq:alg:simple:evd}
      \Cym = \sum_{i=1}^n\lambda_i\vhat_i\vhat_i^\top, \,\,\,\,\, \lambda_1\geq\lambda_2\geq\lambda_3\geq\ldots\geq 0
    \end{equation}
    \STATE Estimate eigenvector centrality: $\uhat=\vhat_1$
    \STATE Sign-correction: $\uhat\leftarrow\uhat\sgn\paren{\abrack{\uhat,\bbone}}$\label{alg:simple:sign-correction}
    \STATE {\bf Output:} The weak ordering $\relates$ induced by $\uhat$
  \end{algorithmic}
\end{algorithm}

In this section, we introduce the main idea of our algorithm with a simple approach.
Based on the discussion in~\cref{sec:problem:ranking}, we consider how well one can do by simply taking the ordering induced by the leading eigenvector of the sample covariance matrix $\Cym$ at face value.
To this end, we propose~\cref{alg:simple},
where the weak ordering $\relates$ induced by the leading eigenvector of $\Cym$ is used to estimate the true eigenvector centrality ranking $\relatesstar$.
Additionally, there is a sign-correction step to resolve the direction of $\uhat$ to ensure, under certain conditions, that it positively correlates with $\u$.  

\subsection{Asymptotic performance}\label{sec:warmup:asymptotic}

We begin by characterizing the \emph{asymptotic} behavior of~\cref{alg:simple}.
That is, we establish the consistency of the simple ranking algorithm as the number of samples $m\to\infty$.
\begin{theorem}\label{thm:simple-asymptotic}
  As $m\to\infty$, \cref{alg:simple} yields an ordering $\relates$ that matches that induced by the true eigenvector centrality $\relatesstar$ with probability $1$.
\end{theorem}
\begin{IEEEproof}
  First, we establish the convergence of $\Cym\to\Cy$ as $m\to\infty$ leveraging the following lemma.
\begin{lemma}[Convergence of $\Cym$~{\cite[Corollary 5.52]{vershynin2010introduction}}]\label{prop:convergence}
  If~\cref{ass:bounded-signals} holds for a collection of signals $\brack{\yl}_{\ell=1}^m$ observed according to~\eqref{eq:system-model}, their sample covariance matrix $\Cym$ satisfies the following with probability at least $1-\eta$:
  \begin{equation}\label{eq:convergence}
    \norm{\Cym - \Cy}_2 < C_0\sqrt{-\log\Paren{\eta}\frac{r}{m}},
  \end{equation}
  where $C_0\in\Theta\paren{\norm{\Cy}_2}$.
\end{lemma}
By \cref{prop:convergence}, $\Cym\to\Cy$ with probability $1$ as $m\to\infty$.
Additionally, under \cref{ass:filter-invertability}, the leading eigenvector of $\Cy$ is unique up to sign, taking value $\u$.
Thus, the leading eigenvector of $\Cym$ converges to that of $\Cy$, \ie $\uhat\to\u$ up to sign.
Additionally, since $\u\succeq 0$, it always holds that $\sgn{\abrack{\u,\bbone}}=1$ \cf{step 5 in Algorithm~\ref{alg:simple}}.
Therefore, in the asymptotic regime, the sign-corrected centrality estimate satisfies $\uhat=\u$.
In this setting, $\relates$ is equivalent to $\relatesstar$.
\end{IEEEproof}

This result is not surprising, given the convergence of the sample covariance to the true covariance as $m\to\infty$.
In practice, though, we only have a finite number of samples and must rely on a noisy covariance estimate, as analyzed next.

\subsection{Non-asymptotic performance}\label{sec:warmup:nonasymptotic}

We now characterize the finite sampling requirements of~\cref{alg:simple} in solving \emph{pairwise} instances of~\cref{prob:ranking}.
That is, we provide a sampling condition under which $\relates$ and $\relatesstar$ are concordant with respect to nodes $i,j$.
\begin{theorem}\label{thm:simple-nonasymptotic}
  If, for some constant $C\in\Theta\paren{\norm{\Cy}_2}$,
  \begin{equation}\label{eq:simple-ranking-requirement}
    m>-\log\Paren{\eta}\frac{C^2}{\delta^2}\max\Brack{\frac{2}{\Paren{u_j-u_i}^2}, \frac{1}{\Abrack{\u,\bbone/\sqrt{n}}^2}},
  \end{equation}
  where $\delta\deq\lambda_1-\lambda_2$ is the eigengap of the covariance matrix $\Cy$, then the ordering $\relates$ yielded by~\cref{alg:simple} and $\relatesstar$ are concordant with respect to $i,j$ with probability at least $1-\eta$.
\end{theorem}
\begin{IEEEproof}
  Without loss of generality, assume $u_j>u_i$.
Then for some perturbation $\beps$ such that $\uhat=\alpha\u+\beps, \norm{\uhat}=1, \alpha\geq 0$, and  $\beps\perp\u$, nodes $i,j$ will be incorrectly ordered by $\relates$ if $\epsilon_j-\epsilon_i<\alpha\paren{u_i-u_j}$.
Conversely, if $\epsilon_j-\epsilon_i$ is lower bounded by $\alpha\paren{u_i-u_j}$, nodes $i,j$ are guaranteed to be correctly ordered by $\relates$.

First, we establish the conditions under which the estimate $\uhat$ yielded by~\cref{alg:simple} guarantees $\alpha\geq0$ in this setting.
\begin{lemma}[Sign-correction]\label{prop:sign-correction}
  \Cref{alg:simple} guarantees that $\abrack{\u,\uhat}\geq 0$ if
  \begin{equation}\label{eq:sign-correction-inequality}
    \abs{\Abrack{\u,\uhat}}>\sqrt{1-\Abrack{\u,\bbone/\sqrt{n}}^2}.
  \end{equation}
\end{lemma}
We leave the proof of this result to \cref{app:sign-correction}.
For convenience, we note that~\eqref{eq:sign-correction-inequality} is equivalent to
\begin{equation}\label{eq:sign-sine-correction-inequality}
  \sin^2\theta\Paren{\u,\uhat}<\Abrack{\u,\bbone/\sqrt{n}}^2.
\end{equation}

Proceeding under the assumption that~\eqref{eq:sign-correction-inequality} in~\Cref{prop:sign-correction} is satisfied, and seeking to minimize $\epsilon_j-\epsilon_i$, consider the following optimization program, for some $\u\succeq 0$:
\begin{mini}
  {\beps}{\epsilon_j-\epsilon_i}
  {\label{eq:simple-proof-program}}{}
  \addConstraint{\norm{\beps}_2^2=1-\alpha^2}
  \addConstraint{\beps\perp\u}
\end{mini}
{From the KKT optimality conditions of~\eqref{eq:simple-proof-program}, it follows that the minimum value of the objective is } $-\sqrt{\paren{1-\alpha^2}\paren{2-\sbrack{u_i-u_j}^2}}$.
That is, if
\begin{equation}\label{eq:simple-proof-precondition}
\alpha^2\Paren{u_i-u_j}^2>\Paren{1-\alpha^2}\Paren{2-\Sbrack{u_i-u_j}^2},
\end{equation}
$\relates$ will correctly order nodes $i,j$ {since this would ensure that $\epsilon_j-\epsilon_i$ is lower bounded by $\alpha\paren{u_i-u_j}$.}

Observing that $\sqrt{1-\alpha^2}=\sin\theta\paren{\u,\uhat}$, this condition is equivalent to the following:
\begin{equation}\label{eq:simple-proof-condition}
  \frac{1-\sin^2\theta\Paren{\u,\uhat}}{\sin^2\theta\Paren{\u,\uhat}}>
  \frac{2}{\Paren{u_i-u_j}^2}-1.
\end{equation}

The convergence of $\sin\theta\paren{\u,\uhat}\to 0$ with $m$ follows from \cref{prop:convergence} and~\cite[Theorem 2]{yu2014daviskahanstat}, stated in the following proposition:
\begin{lemma}[Angle between $\uhat$ and $\u$]\label{prop:alignment}
  Under the same conditions as~\cref{prop:convergence}, for the leading eigenvectors $\v_1,\vhat_1$ of $\Cy, \Cym$, respectively, the following holds with probability at least $1-\eta$:
  \begin{equation}\label{eq:alignment}
    \sin\theta\Paren{\v_1,\vhat_1} < 2\frac{C_0}{\delta}\sqrt{-\log\Paren{\eta}\frac{r}{m}},
  \end{equation}
  where $\delta\deq\lambda_1-\lambda_2$ is the population eigengap.
\end{lemma}
Requiring both~\eqref{eq:simple-proof-condition} and~\eqref{eq:sign-sine-correction-inequality} to hold, applying~\cref{prop:alignment} yields the sampling requirement~\eqref{eq:simple-ranking-requirement}, as desired.
\end{IEEEproof}

\Cref{thm:simple-nonasymptotic} characterizes the sampling requirements to ensure both that the orientation of $\uhat$ is the correct one (\ie $\abrack{\u,\uhat}>0$) and that the perturbation ${\beps}$ is sufficiently small for \emph{a single} pairwise ranking problem to be determined correctly.
Indeed, the $\max\paren{\cdot}$ term in~\eqref{eq:simple-ranking-requirement} reflects the need for both conditions to hold. Notice that the first argument depends on the specific nodes to be ranked whereas the second argument encodes a global measurement of the difficulty of recovering the ranking {as dictated by how close $\u$ is to the constant vector, \ie $\abrack{\u,\bbone/\sqrt{n}}$}.

Once the sampling condition for sign-correction is met, \eqref{eq:simple-ranking-requirement} is determined by the difference in centralities between the two nodes in question.
To further shed light on this, we notice that in the program~\eqref{eq:simple-proof-program}, the $\beps$ attaining the described minimum is
\begin{equation}\label{eq:simple-minimizing-error}
  \beps = \sqrt{\frac{1-\alpha^2}{2-\Paren{u_j-u_i}^2}}\Paren{\e_i-\e_j+\Sbrack{u_j-u_i}\u}.
\end{equation}
Recalling the assumption that $u_j>u_i$, the orthogonal error $\beps$ between $\uhat$ and $\u$ will preserve the ordering of \emph{all pairs of nodes} that do not include $i$ or $j$.
That is, the worst-case perturbation for the pairwise ranking problem between two nodes $i,j$ is ideal for almost every other pairwise ranking problem.
So, if one were to consider the number of samples required to correctly rank some \emph{fixed proportion} of pairwise ranking problems at some resolution, the derived requirement would be less than that put forth in \cref{thm:simple-nonasymptotic}, if only by a constant factor.
We demonstrate this in~\cref{sec:experiments:erdos-renyi}.
{Additionally, if $u_j=u_i$, then the sampling requirement of \cref{thm:simple-nonasymptotic} is undefined, so we only guarantee concordant ordering in the asymptotic setting of \cref{thm:simple-asymptotic}.}

\begin{remark}[Other perturbation bounds]\label{remark:simple}
  \Cref{thm:simple-nonasymptotic} provides the number of samples required to guarantee a sufficiently small elementwise perturbation.
  This sampling requirement was derived via classical perturbation bounds~\cite{daviskahan1970}, where the alignment of eigenvectors is characterized in terms of the angle between them.
  In this way, we bound the perturbation between pairs of elements in the leading eigenvector.

  Several existing works directly provide uniform, elementwise bounds on eigenvector perturbation~\cite{fan2018perturbation,eldridge2018}.
  However, these results typically depend on incoherence of the eigenspace, or assumptions on the randomness of the noise, \eg \iid\ Gaussian noise~\cite{koltchinskii2016perturbation}.
  In contrast, \cref{thm:simple-nonasymptotic} does \emph{not} directly depend on properties such as incoherence or delocalization, thus holding more generally.
  For instance, the eigenvector centralities of \WS random graphs have been demonstrated to be highly localized~\cite{farkas2001spectra}.
  Such structures do not satisfy the requirements for strong perturbation bounds in the $\ell_\infty$-norm in~\cite{fan2018perturbation}.
  Additionally, the error between a sample covariance matrix and the population covariance matrix of a multivariate Gaussian does not have appealing properties such as having \iid\ Gaussian entries, as required in~\cite{koltchinskii2016perturbation}.
\end{remark}

\section{Ranking with Confidence}\label{sec:ranking-prior}

\begin{algorithm}[tb]
  \caption{Blind centrality ranking with thresholding}\label{alg:threshold-ranking}
  \begin{algorithmic}[1]
    \STATE {\bf Input:} Graph signals $\brack{\yl}_{\ell=1}^m$, threshold $\tau>0$
    \STATE {\bf Returns:} Partial ordering $\relates_\tau$
    \STATE Compute the sample covariance matrix $\Cym$~\cf{\eqref{eq:alg:simple:covariance}}
    \STATE Compute the eigenvector decomposition of $\Cym$, $\brack{\vhat_i,\lambda_i}_{i=1}^n$~\cf{\eqref{eq:alg:simple:evd}}
    \STATE Estimate eigenvector centrality: $\uhat=\vhat_1$
    \STATE Sign-correction: $\uhat\leftarrow\uhat\sgn\paren{\abrack{\uhat,\bbone}}$
    \FOR{$i,j\in\Verts\times\Verts$}
    \IF{$\abs{\widehat{u}_i-\widehat{u}_j}\leq\tau$}
    \STATE $\relates_\tau$ abstains from ordering $i,j$
    \ELSIF{$\widehat{u}_i-\widehat{u}_j>\tau$}
    \STATE Set $j\relates_\tau i$
    \ELSE
    \STATE Set $i\relates_\tau j$
    \ENDIF
    \ENDFOR
    \STATE {\bf Output:} Partial ordering $\relates_\tau$
  \end{algorithmic}
\end{algorithm}

After establishing the performance of \cref{alg:simple}, a simple approach to~\cref{prob:ranking}, we now present a refined algorithm that allows one to incorporate \emph{confidence} in the estimate of the eigenvector centrality based on sampling noise.

In~\cref{alg:simple}, the ordering induced by $\uhat$ is taken at face value.
That is, if $\uhat_i>\uhat_j$, the conclusion is drawn that $\u_i>\u_j$.
However, this does not distinguish between cases where $\abs{\uhat_i-\uhat_j}$ is very large and cases where $\uhat_i\approx\uhat_j$.
In this context, one could use a threshold to measure the significance of an estimated ordering.
That is, for some $\tau>0$, if $\abs{\uhat_i-\uhat_j}>\tau$, we make the claim that $\uhat$ correctly orders $i$ and $j$.
Conversely, if $\abs{\uhat_i-\uhat_j}\leq\tau$, we abstain from ordering nodes $i,j$ based on $\uhat$.
In this way, by abstaining from inducing an order relation on nodes when their difference in $\uhat$ is too small, $\Verts$ is endowed with a \emph{partial ordering}.
Motivated by this, we propose~\cref{alg:threshold-ranking}, where a threshold $\tau$ is introduced to qualify the ordering $\relates$ induced by the leading eigenvector of $\Cym$.
That is, a difference in centralities $\abs{\uhat_i-\uhat_j}$ is required to be sufficiently large in order for the partial ordering to not abstain from making a comparison.

Notice that~\cref{alg:simple} is indeed a special case of~\cref{alg:threshold-ranking} where $\tau=0$.
That is, \cref{alg:simple} can be interpreted as having \emph{full confidence} in the ordering induced by the estimated centrality $\uhat$.
The remaining question, then, relates to how to choose the threshold $\tau$ for~\cref{alg:threshold-ranking}.
We begin by defining a \emph{viable} threshold $\tau$.
\begin{defn}[Viable threshold]\label{defn:viable}
  For an estimated eigenvector centrality $\uhat\in\Real^n$ compared to a true eigenvector centrality $\u\in\Real^n$, a threshold $\tau>0$ is said to be \emph{viable} if for any $i,j\in\sbrack{n}$ where $\widehat{u}_j-\widehat{u}_i>\tau$, it also holds that $u_j>u_i$.
\end{defn}
By~\cref{defn:viable}, it is immediately clear that if~\cref{alg:threshold-ranking} is run with a viable threshold $\tau$, the resulting partial ordering $\relates_\tau$ will either be concordant with $\relatesstar$ or abstain from ranking each pair of nodes.
It is also true that for some viable threshold $\tau_0$, any $\tau_1>\tau_0$ is also a viable threshold.
However, choosing a larger threshold for~\cref{alg:threshold-ranking} is clearly not desirable, as it would lead to needless abstention from otherwise correct pairwise orderings.

\subsection{Convergence of the minimum viable threshold with $m$}\label{sec:ranking-prior:sampling}

Similarly to the proof of~\cref{thm:simple-nonasymptotic}, we can characterize the convergence of $\uhat$ to $\u$ as the number of samples $m$ grows, measured by the inner product $\abrack{\u,\uhat}=\alpha$.
Noting that $\sqrt{1-\alpha^2}=\sin\theta\paren{\u,\uhat}$, the following proposition holds due to~\cref{prop:alignment}.

\begin{prop}\label{prop:threshold-sample-bound}
  There exists a constant $C\in\Theta\paren{\norm{\Cy}_2}$ such that for $0<\eta\leq 1$, $\tau>0$, if
  \begin{equation}\label{eq:threshold-sample-bound}
    \frac{C}{\delta}\sqrt{\frac{-\log\Paren{\eta}}{m}}<\min\Brack{\frac{\tau}{\sqrt{2}},\Abrack{\u,\bbone/\sqrt{n}}},
  \end{equation}
  then $\tau$ is a viable threshold with probability at least $1-\eta$.
\end{prop}
\begin{IEEEproof}
  Similar to the proof of~\Cref{thm:simple-nonasymptotic}, we seek to establish conditions under which $m$ is large enough to guarantee $\alpha=\abrack{\u,\uhat}>0$, as well as guaranteeing that the perturbation $\beps$, where $\uhat=\alpha\u+\beps$ is small enough, to ensure that $\tau$ is a viable threshold.

By~\Cref{prop:sign-correction}, $\alpha>0$ is guaranteed by~\Cref{alg:threshold-ranking} when $\abs{\abrack{\u,\uhat}}\geq\sqrt{1-\abrack{\u,\bbone/\sqrt{n}}^2}$.

From {the discussion following the proof of \cref{thm:simple-nonasymptotic}~\cf{\eqref{eq:simple-minimizing-error}}}, for $\alpha>0$ and $\norm{\u}=\norm{\uhat}=1$, the following holds:
\begin{equation}
  {\inf_{\u,\uhat\colon\Abrack{\u,\uhat}=\alpha,u_j> u_i}}\widehat{u}_j-\widehat{u}_i=-\sqrt{2\Paren{1-\alpha^2}}.
\end{equation}
Thus, setting $\tau=\sqrt{2\Paren{1-\alpha^2}}$ is guaranteed to abstain from all incorrect pairwise orderings induced by $\uhat$.
Equivalently, for some chosen $\tau>0$ to be guaranteed viable, we require $\alpha>\sqrt{1-\frac{\tau^2}{2}}$.

Finally, for both conditions to be met ($\alpha >0$ and viability of the threshold), we require $m$ to be such that both $\alpha^2>1-\abrack{\u,\bbone/\sqrt{n}}^2$ and $\alpha^2>1-\frac{\tau^2}{2}$.
This can be more concisely written as
\begin{equation}\label{eq:threshold-alpha-requirement}
  \sin^2\theta\Paren{\u,\uhat}<\min\Brack{\frac{\tau^2}{2},\Abrack{\u,\bbone/\sqrt{n}}^2}.
\end{equation}
Applying~\Cref{prop:alignment} to~\eqref{eq:threshold-alpha-requirement} yields~\eqref{eq:threshold-sample-bound}, as desired.
\end{IEEEproof}

Effectively, \cref{prop:threshold-sample-bound} prescribes choosing
\begin{equation}\label{eq:threshold-prescription}
  \tau=\frac{C}{\sqrt{m}},
\end{equation}
for some $C>0$, when $m$ is sufficiently large.
Setting $\tau\sim\frac{1}{\sqrt{m}}$ has the appealing property of converging to $0$ as $m\to\infty$, which is the minimum viable threshold when $\uhat=\u$~\cf{\cref{thm:simple-asymptotic}}.
In \cref{sec:experiments}, we demonstrate that this functional relationship between $m$ and the minimum viable threshold $\tau$ is tight in practice.

For a fixed number of samples $m$, \eqref{eq:threshold-sample-bound} dictates, via $\abrack{\u,\bbone/\sqrt{n}}$, a strict lower bound $\eta_0$ of the failure probability $\eta$ that can be attained, \ie $\eta_0<\eta\leq 1$.
Then, as $\eta\to\eta_0^+$, the minimum threshold $\tau$ that satisfies~\eqref{eq:threshold-sample-bound} increases.
This is intuitive, as a larger threshold has a higher probability of being viable for any given centrality estimate $\uhat$.

\begin{remark}[Enhancing sign-correction]
  In practice, one could employ prior knowledge on the structure of $\u$ to enhance the sign-correction step for $\uhat$, by replacing $\bbone/\sqrt{n}$ with a sufficiently close approximation of the eigenvector centrality $\utilde$.
  Then, the limitation on $\eta$ determined by $\abrack{\u,\bbone/\sqrt{n}}$ would be relaxed.
  That is, \eqref{eq:threshold-sample-bound} would be replaced by the condition
  \begin{equation}
    \frac{C}{\delta}\sqrt{\frac{-\log\Paren{\eta}}{m}}<\min\Brack{\frac{\tau}{\sqrt{2}},\Abrack{\u,\utilde}}.
  \end{equation}
  Reasonable approximations $\utilde$ could include the vector of node degrees, or the eigenvector centrality induced by some auxiliary network construction.
\end{remark}

\section{Numerical Experiments}\label{sec:experiments}

We illustrate the behavior of~\cref{alg:simple,alg:threshold-ranking} via numerical experiments on synthetic and real-world graphs.
We begin by demonstrating the relationship between sample size and the performance of both algorithms on an \ER random graph. Then, we proceed to investigate the influence of the underlying graph structure on the difficulty of~\cref{prob:ranking} via simulations on a graphon-based model.
We conclude with a centrality analysis based on U.S. Senate voting records compared to an existing model for interpreting voting patterns.

\subsection{Sampling requirements of \Cref{prob:ranking}}\label{sec:experiments:erdos-renyi}

\begin{figure*}[t]
  \centering
  \resizebox{\textwidth}{!}{\begin{tikzpicture}

  \begin{groupplot}[
    group style={
      group name=myplots,
      group size=4 by 1,
      horizontal sep=2cm},
    label style={font=\Large},
    every tick label/.append style={font=\Large},
    legend style={font=\Large,draw=white!80.0!black},
    tick align=inside,
    tick pos=both,
    x grid style={gray!30},
    xmajorgrids,
    xminorgrids,
    xtick style={color=black},
    xticklabel style={
      /pgf/number format/fixed,
      /pgf/number format/precision=2
    },
    y grid style={gray!30},
    ymajorgrids,
    yminorgrids,
    ytick style={color=black},
    yticklabel style={
      /pgf/number format/fixed,
      /pgf/number format/precision=2
    },
    scaled y ticks=false,
    width=0.4\textwidth,
    height=0.4\textwidth,
    scale only axis,
    ]

    \nextgroupplot[
      legend pos={north west},
      legend cell align={left},
      xlabel={Node index},
      ylabel={Eigenvector centrality},
      no markers,
      xmin=0, xmax=100,
      ymin=0, ymax=0.27,
    ]
    \addplot table[x=idx, y=u] {./outdata/er-threshold-demo/centrality.csv};

    \nextgroupplot[
    xlabel={$\abs{u_i-u_{n/2}}$},
    ylabel={Ranking error rate $\eta$},
    xmode=normal,
    ymode=log,
    legend columns=1,
    legend cell align=center,
    legend pos={south east},
    ymin=0.01, ymax=1,
    ]
    \addplot+[only marks] table[x expr=\thisrow{difficulty}, y expr=1-\thisrow{p100}] {./outdata/er-threshold-demo/comparison.csv};
    \addlegendentry{$m=10^2$}
    \addplot+[only marks] table[x expr=\thisrow{difficulty}, y expr=1-\thisrow{p1000}] {./outdata/er-threshold-demo/comparison.csv};
    \addlegendentry{$m=10^3$}
    \addplot+[only marks] table[x expr=\thisrow{difficulty}, y expr=1-\thisrow{p10000}] {./outdata/er-threshold-demo/comparison.csv};
    \addlegendentry{$m=10^4$}
    \addplot+[only marks] table[x expr=\thisrow{difficulty}, y expr=1-\thisrow{p100000}] {./outdata/er-threshold-demo/comparison.csv};
    \addlegendentry{$m=10^5$}

    \addplot+[no marks, solid, black, thin, domain=0:0.2] {10^(-0.08487*100*x^2-0.48)};
    \addplot+[no marks, solid, black, thin, domain=0:0.2] {10^(-0.08487*1000*x^2-0.48)};
    \addplot+[no marks, solid, black, thin, domain=0:0.05] {10^(-0.08487*10000*x^2-0.48)};
    \addplot+[no marks, solid, black, thin, domain=0:0.02] {10^(-0.08487*100000*x^2-0.48)};

    \nextgroupplot[
    separate axis lines,
    xlabel={Sample size $m$},
    ylabel={Minimum viable threshold},
    xmode=log,
    ymode=normal,
    legend cell align=center,
    legend pos=north east,
    xmin=99, xmax=100001,
    ymin=0.0, ymax=0.5,
    scale only axis,
    axis y line*=left,
    axis x line*=box,
    every outer y axis line/.append style={red},
    ylabel style={red},
    y tick style={red},
    y tick label style={red},
    ]
    \addplot+[only marks, mark size=3] table[x=m, y=threshold] {./outdata/er-threshold-demo/thresholds.csv}; \label{minthresh-points}
    \addplot+[no marks, red, solid, thin, domain=100:100000] {4.440/sqrt(x)}; \label{minthresh-fit}
    
    \nextgroupplot[
    xlabel={Sample size $m$},
    ylabel={Spearman's $\rho$},
    xmode=log,
    ymode=normal,
    legend cell align=center,
    legend pos=south east,
    ymin=0.0, ymax=1.0,
    yticklabel pos=right,
    ]
    \addplot+[mark size=3] table[x=m, y=blind] {./outdata/er-threshold-demo/nti-comparison.csv};
    \addlegendentry{Blind}
    \addplot+[mark size=4] table[x=m, y=kalofolias] {./outdata/er-threshold-demo/nti-comparison.csv};
    \addlegendentry{Kalofolias}
    \addplot+[mark size=3] table[x=m, y=spectemp] {./outdata/er-threshold-demo/nti-comparison.csv};
    \addlegendentry{Spec. Temp.}
    
  \end{groupplot}

  \begin{groupplot}[
    group style={
      group name=myplots_overlay,
      group size=4 by 1,
      horizontal sep=2cm},
    label style={font=\Large},
    every tick label/.append style={font=\Large},
    legend style={font=\Large,draw=white!80.0!black},
    tick align=inside,
    tick pos=both,
    x grid style={gray!30},
    xmajorgrids,
    xminorgrids,
    xtick style={color=black},
    xticklabel style={
      /pgf/number format/fixed,
      /pgf/number format/precision=2
    },
    y grid style={gray!30},
    ymajorgrids,
    yminorgrids,
    ytick style={color=black},
    yticklabel style={
      /pgf/number format/fixed,
      /pgf/number format/precision=2
    },
    scaled y ticks=false,
    width=0.4\textwidth,
    height=0.4\textwidth,
    scale only axis,
    ]

    \nextgroupplot[group/empty plot]

    \nextgroupplot[group/empty plot]
  
    \nextgroupplot[
    xlabel={},
    ylabel={Completeness},
    xmode=log,
    ymode=normal,
    legend cell align=center,
    legend style={
      at={(0.5,0.99)},
      anchor=north,
      minimum width=2cm
    },
    xmin=99, xmax=100001,
    ymin=0.0, ymax=1.0,
    scale only axis,
    axis y line*=right,
    axis x line*=box,
    hide x axis,
    xmajorgrids=false, xminorgrids=false,
    ymajorgrids=false, yminorgrids=false,
    every outer y axis line/.append style={blue},
    ylabel style={blue},
    y tick style={blue},
    y tick label style={blue},
    ]
    \addlegendimage{/pgfplots/refstyle=minthresh-points}\addlegendentry{Min. Viable $\tau$}
    \addlegendimage{/pgfplots/refstyle=minthresh-fit}\addlegendentry{$C/\sqrt{m}$}
    \pgfplotsset{cycle list shift=1}
    \addplot+[mark size=4] table[x=m, y=completeness] {./outdata/er-threshold-demo/thresholds.csv};
    \addlegendentry{Completeness}

    \nextgroupplot[group/empty plot]
P    
  \end{groupplot}
  
  \node[below left,fill=white] at (myplots c1r1.north east) {\LARGE\textbf{A}};
  \node[below left,fill=white] at (myplots c2r1.north east) {\LARGE\textbf{B}};
  \node[below left,fill=white] at (myplots c3r1.north east) {\LARGE\textbf{C}};
  \node[below left,fill=white] at (myplots c4r1.north east) {\LARGE\textbf{D}};

\end{tikzpicture}

}
  \caption{
    Ranking algorithm on \ER random graph.
    (A) Eigenvector centrality of a drawn \ER graph with $n=100$ and $p=\frac{\log{n}}{n}$.
    (B) Ranking error rate for every node relative to node $50$, for $m\in\brack{10^2,\ldots,10^5}$ plotted against difference in centrality.
    {(C, left axis)} Minimum viable threshold $\tau$ against sample size $m$, compared to best fit curve $C/\sqrt{m}$.
    {(C, right axis)} Completeness of partial ordering with minimum viable threshold $\tau$ against sample size $m$.
    {(D) Spearman correlation of eigenvector centrality estimate against sample size $m$ for network topology inference techniques (`Kalofolias'~\cite{kalofolias2016smooth} and `Spec. Temp.'~\cite{segarra2017topo}) and the proposed blind method.}
  }
  \label{fig:experiments:erdos-renyi}
\end{figure*}
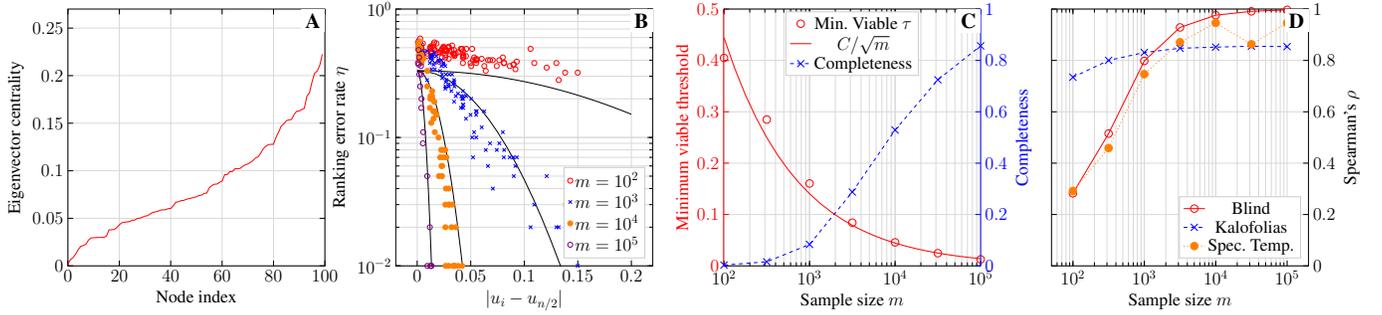

To demonstrate the relationship between the number of samples $m$ and the performance of~\cref{alg:simple,alg:threshold-ranking}, we consider an \ER graph with $n=100$ and $p={\log\paren{n}}/{n}$.
Due to the sparsity level of this graph, there is a noticeable structure in the eigenvector centrality profile -- illustrated in~\letterref{fig:experiments:erdos-renyi}{A} -- where, \eg the most central node attains a centrality value $3.34$ times larger than the average centrality in the graph.

The excitation signals $\wl$ are drawn from a white, Gaussian distribution, and the graph filter $\cH$ takes the square root of the absolute value of the graph's eigenvalues.
That is, the filter $\cH\paren{\lambda}$ in~\eqref{eq:graph-filter} is given by $\cH\paren{\lambda} = \sqrt{\abs{\lambda}}$, so that the covariance matrix $\Cy$ satisfies $\Cy = \sum_{l=0}^n\abs{\lambda_l}\v_l\v_l^\top$, where $\lambda_l,\v_l$ are the eigenvalues and eigenvectors of $\A$.

The range of centralities in $\u$ yields a wide diversity of difficulties across pairwise instances of~\cref{prob:ranking}.
Nodes with similar centralities have high sampling requirements by~\cref{thm:simple-nonasymptotic} and nodes with large differences in centrality have low sampling requirements.
To see the effect of the number of samples in this scenario, we pick the node with the median centrality (index $50$) as a reference, and evaluate the probability of correctly (or incorrectly) ordering each node against the reference node for multiple sample sizes across $100$ trials.
The outcome is shown in~\letterref{fig:experiments:erdos-renyi}{B}, where we plot the probability of incorrect ranking as a function of $|u_i-u_{n/2}|$.

{In addition to plotting the ranking error rates for each node, we consider how well these probabilities are explained by~\cref{thm:simple-nonasymptotic}.
To do this, we find constants $C_0, C_1$ that best fit the expression
\begin{equation}\label{eq:probability-regression}
  \log\Paren{\eta_{i,m}}=C_0m\Paren{u_i-u_{n/2}}^2+C_1
\end{equation}
over all nodes $i$ and sample sizes $m$, where $\eta_i$ is the empirical error rate for node $i$.
Although the bound put forth in~\cref{thm:simple-nonasymptotic} would imply $C_1=0$, we include it as a free parameter here in line with the discussion following the proof of~\cref{thm:simple-nonasymptotic}.

The results of this regression task are plotted in~\letterref{fig:experiments:erdos-renyi}{B} in black for each sample size $m$.
It becomes evident that the functional form of the error rates for each node is matched fairly well by the curve~\eqref{eq:probability-regression}, especially for larger sample sizes.}

Additionally, in evaluating~\cref{alg:threshold-ranking}, we consider the convergence of the minimum viable threshold $\tau$ with $m$.
As stated in~\cref{prop:threshold-sample-bound}, the minimum viable threshold is expected to scale according to ${\cal O}\paren{1/\sqrt{m}}$.
In~\letterref{fig:experiments:erdos-renyi}{C}, the average minimum viable threshold over $100$ trials is plotted against the sample size $m$, alongside the best-fit inverse square-root curve.
One can see that the empirical minimum viable threshold is indeed well-approximated by $C/\sqrt{m}$.
That is, the upper bound on the minimum viable threshold in~\cref{prop:threshold-sample-bound} appears to be tight in this setting.

Also shown in~\letterref{fig:experiments:erdos-renyi}{C} is the average \emph{completeness} of the partial order returned by~\cref{alg:threshold-ranking} using the minimum viable threshold.
We define the completeness of a partial ordering $\relates$ as the number of pairs of nodes that $\relates$ does not abstain from comparing, divided by $\binom{n}{2}$.
Putting it differently, the completeness is equal to $1$ minus the rate of abstention.
Since the minimum viable threshold for a partial order $\relates$ guarantees no discordance with $\relatesstar$ for all pairs of nodes, it must reject a larger proportion of pairwise ordering induced by $\uhat$ when the threshold is large.
It is apparent that as the minimum viable threshold decreases with $m$, the completeness of the associated partial order approaches $1$.

{Finally, we compare our `blind' inference approach to network topology inference techniques.
  Recalling that the goal of our proposed method is to bypass the complex and data-hungry network topology inference techniques, we compare how well these methods perform in extracting a final estimate of the eigenvector centrality.
  In~\letterref{fig:experiments:erdos-renyi}{D}, we plot the average Spearman correlation~\cite{cordernonparam} of the blind eigenvector centrality estimate for multiple sample sizes, as well as the same quantity for the leading eigenvector of adjacency matrices inferred using existing network topology inference approaches.
  It is apparent that for a small number of samples, \cite{kalofolias2016smooth} outperforms other methods, including ours.
  This can be explained by the {sparsity regularization}: although the approach taken by~\cite{kalofolias2016smooth} does not explicitly use eigenvector computations, the sparsity enforced by the choice of hyperparameters imparts additional knowledge of the graph structure.
  Thus, the variance in $\uhat$ brought on by a small number of samples is traded for bias by the sparsity regularization in~\cite{kalofolias2016smooth}.
  As the number of samples increases, however, this bias-variance tradeoff is no longer advantageous, and the unbiased estimate of the eigenvector centrality performs better.
  Additionally, the performance of the spectral templates approach proposed in~\cite{segarra2017topo} is upper bounded by that of our proposed method.
Although it also seeks sparse graphs explained by the data, it directly leverages the eigenvectors of $\Cym$, and does not have a significant advantage for the purpose of centrality ranking.}

\subsection{Influence of graph structure on~\Cref{prob:ranking}}\label{sec:experiments:circular-ranking}

\begin{figure*}[t]
  \centering
  \resizebox{\textwidth}{!}{\begin{tikzpicture}

  \begin{groupplot}[
    group style={
      group name=myplots,
      group size=5 by 1,
      horizontal sep=1.75cm},
    label style={font=\large},
    every tick label/.append style={font=\large},
    legend style={font=\large,draw=white!80.0!black},
    tick align=inside,
    tick pos=both,
    x grid style={gray!30},
    xmajorgrids,
    xminorgrids,
    xtick style={color=black},
    xticklabel style={
      /pgf/number format/fixed,
      /pgf/number format/precision=2,
    },
    y grid style={gray!30},
    ymajorgrids,
    yminorgrids,
    ytick style={color=black},
    yticklabel style={
      /pgf/number format/fixed,
      /pgf/number format/precision=2,
    },
    scaled y ticks=false,
    width=0.4\textwidth,
    height=0.4\textwidth
    ]

    \nextgroupplot[
      legend pos={north west},
      legend cell align={left},
      xlabel={Node index},
      ylabel={Eigenvector centrality},
      no markers,
      xmin=0, xmax=250,
      ymin=0.04, ymax=0.1,
    ]
    \pgfplotsset{cycle list shift=-1}
    \addplot table[x=idx, y=u_mean] {./outdata/circthresh-wet08-hard/centrality.csv};
    \addlegendentry{$\gamma=0.8$}
    \addplot table[x=idx, y=u_mean] {./outdata/circthresh-wet02-hard/centrality.csv};
    \addlegendentry{$\gamma=0.2$}
    \addplot+[black, solid, thin, domain=0:250] {(sqrt(3/(3+sqrt(5)))*(sqrt(10)*(x/250)^2/2+1/sqrt(2))+0.5687)/24.38};
    \addplot+[black, solid, thin, domain=0:250] {(sqrt(3/(3+sqrt(5)))*(sqrt(10)*(x/250)^2/2+1/sqrt(2))+9.1)/158.72};

    \nextgroupplot[
    tick align=outside,
    point meta min=0.48,
    point meta max=1.0,
    colorbar horizontal,
    colorbar style={
      at={(0.5,1.03)},
      anchor=south,
      xticklabel pos=upper
    },
    colormap={WB}{color=(black) color=(white)},
    xlabel={Node index},
    ylabel={Node index},
    xmin=0, xmax=250,
    ymin=0, ymax=250,
    ]
    \addplot graphics [xmin=0, xmax=250, ymin=0, ymax=250] {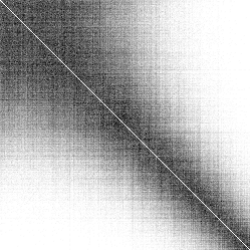};

    \nextgroupplot[
    tick align=outside,
    point meta min=400,
    point meta max=10000,
    colorbar horizontal,
    colorbar style={
      at={(0.5,1.03)},
      anchor=south,
      xticklabel pos=upper
    },
    colormap={WB}{color=(white) color=(black)},
    xlabel={Node index},
    ylabel={Node index},
    xmin=0, xmax=250,
    ymin=0, ymax=250,
    ]
    \addplot graphics [xmin=0, xmax=250, ymin=0, ymax=250] {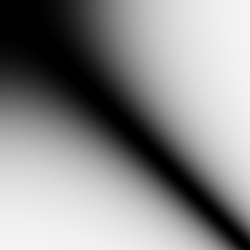};

    \nextgroupplot[
    xlabel={Sample size $m$},
    ylabel={Minimum viable threshold},
    xmode=log,
    ymode=normal,
    legend cell align=center,
    legend style={
      at={(0.97,0.97)},
      anchor=north east,
    },
    ymin=0.0,
    ]
    \addplot+[only marks, mark size=3] table[x=m, y=threshold] {./outdata/circthresh-wet08-hard/thresholds.csv};
    \addlegendentry{$\tau$: $\gamma=0.8$}
    \addplot+[only marks, mark size=3] table[x=m, y=threshold] {./outdata/circthresh-wet02-hard/thresholds.csv};
    \addlegendentry{$\tau$: $\gamma=0.2$}
    \addplot+[no marks, red, solid, thin, domain=250:100000] {1.193/sqrt(x)};
    \addlegendentry{$C_1/\sqrt{m}$: $\gamma=0.8$}
    \addplot+[no marks, blue, solid, thin, domain=250:100000] {0.796/sqrt(x)};
    \addlegendentry{$C_2/\sqrt{m}$: $\gamma=0.2$}

    \nextgroupplot[
    xlabel={Sample size $m$},
    ylabel={Completeness},
    xmode=log,
    ymode=normal,
    legend cell align=center,
    legend pos=north west,
    ymin=0.0, ymax=1.0,
    ]
    \addplot table[x=m, y=completeness] {./outdata/circthresh-wet08-hard/thresholds.csv};
    \addlegendentry{$\gamma=0.8$}
    \addplot table[x=m, y=completeness] {./outdata/circthresh-wet02-hard/thresholds.csv};
    \addlegendentry{$\gamma=0.2$}
  
  \end{groupplot}
  
  \node[below left,fill=white] at (myplots c1r1.north east) {\LARGE\textbf{A}};
  \node[below left,fill=white] at (myplots c2r1.north east) {\LARGE\textbf{B}};
  \node[below left,fill=white] at (myplots c3r1.north east) {\LARGE\textbf{C}};
  \node[below left,fill=white] at (myplots c4r1.north east) {\LARGE\textbf{D}};
  \node[below left,fill=white] at (myplots c5r1.north east) {\LARGE\textbf{E}};
\end{tikzpicture}

}
  \caption{
    Ranking algorithm for the mixed circular random graph model.
    (A) Mean eigenvector centrality of 100 graphs drawn from CRGM with $\gamma \in \{0.2,0.8\}$.
    Variance of centralities for each node are consistently on the order of $10^{-5}$, so are not plotted here.
    Additionally, the expected centralities according to~\cref{claim:crgm-concentration} are plotted in black.
    (B) Empirical probabilities of pairwise ranking success for $\gamma=0.8$ when $m=1000$.
    (C) Expected sampling requirement $\paren{u_i-u_j}^{-2}$ according to \cref{thm:simple-nonasymptotic} and \cref{claim:crgm-concentration} for $\gamma=0.8$.
    (D) Minimum viable thresholds $\tau$ against sample size $m$, compared to best fit curves $C/\sqrt{m}$ for $\gamma \in \{0.2,0.8\}$.
    (E) Completeness of partial ordering with minimum viable threshold $\tau$ against sample size $m$ for $\gamma \in \{0.2,0.8\}$.
  }
  \label{fig:experiments:circular}
\end{figure*}
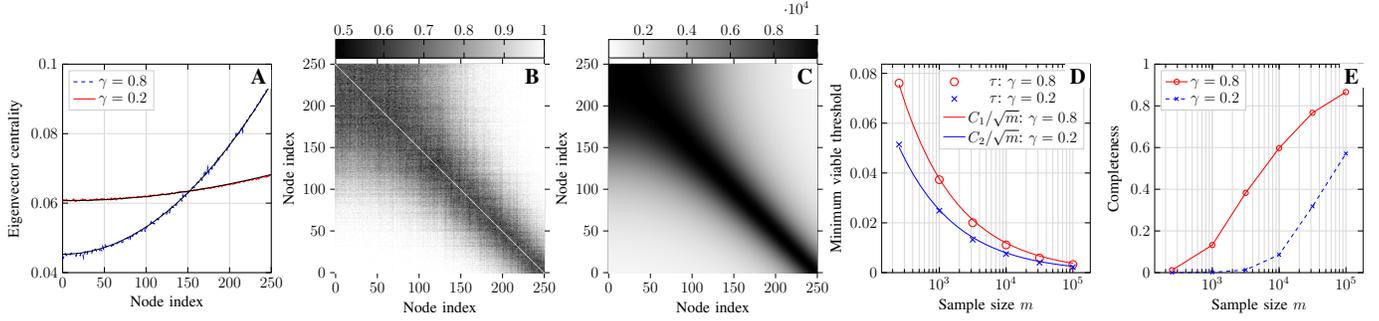

Consider graphs drawn from a \textit{circular random graph model} (CRGM)~\cite{medina2018graphon}.
For a fixed set of $n$ nodes indexed with the integers $\sbrack{n}$, each edge between nodes $i$ and $j$ exists (independently) with probability $\frac{\paren{i/n}^2+\paren{j/n}^2}{2}$.
This model has an increasing centrality structure, with nodes of higher index being more central than nodes of lower index.
As shown in~\cite{medina2018graphon,dasaratha2017centrality}, the eigenvector centrality of a large graph drawn from such a random graph model concentrates around its expectation, which we further validated in our experiments.

To understand how the centrality structure of a graph model influences the difficulty of~\cref{prob:ranking}, we further introduce the \textit{mixed CRGM}.
For some parameter $\gamma\in\sbrack{0,1}$, the probability of existence for the edge $i,j$ is given by $\gamma\frac{\paren{i/n}^2+\paren{j/n}^2}{2}+\paren{1-\gamma}$.
When $\gamma=1$, this model is exactly the CRGM.
As $\gamma$ approaches $0$, the mixed CRGM resembles a fully connected graph, which has a constant centrality structure.
Thus, we expect~\cref{prob:ranking} to be easier for graphs drawn from a model with $\gamma\approx 1$ than those from a model with $\gamma\approx 0$.
We characterize this flattening of the eigenvector centrality in the following claim.
\begin{claim}[Concentration of centrality for mixed CRGM]\label{claim:crgm-concentration}
  For sufficiently large $n$, the following statements hold:
  \begin{enumerate}
  \item The leading eigenvector of graphs drawn from the CRGM concentrate around the vector $\v$, where
    \begin{equation}\label{eq:crgm-centrality}
      v_i = c_0 \Paren{\frac{\sqrt{10}}{2}\Paren{\frac{i}{n}}^2 + \frac{1}{\sqrt{2}}},
    \end{equation}
    with $c_0$ such that $\v$ has unit norm.
  \item The leading eigenvector of graphs drawn from a mixed CRGM with mixture parameter $\gamma > 0$ concentrates around the vector $\widetilde{\v}$, where
    \begin{equation}\label{eq:mixed-crgm-centrality}
      \widetilde{v}_i = c_1\Paren{v_i+\frac{\beta}{\sqrt{n}}},\ \beta \approx 2.275 \frac{1-\gamma}{\gamma},
    \end{equation}
    with $c_1$ such that $\widetilde{\v}$ has unit norm.
  \end{enumerate}
\end{claim}
The first part of~\cref{claim:crgm-concentration} is a rephrasing of~\cite[Example 4.2.1]{medina2018graphon}.
The second part can be shown via a graphon-based argument combining~\cite[Theorem 2]{medina2018graphon} and a rank-one update analysis.

\Cref{claim:crgm-concentration} states that the eigenvector centralities of graphs drawn from the mixed CRGM are close to the leading eigenvector of the \emph{expected} adjacency matrix for the model.
Moreover, it characterizes the \emph{rate} at which the eigenvector centrality of graphs drawn from a mixed CRGM approaches the constant vector as $\gamma\to 0$.
To verify this, we plot the expected eigenvector centrality based on~\eqref{eq:mixed-crgm-centrality} for mixed CRGMs of size $n=250$ with $\gamma \in \{0.2, 0.8\}$ on~\letterref{fig:experiments:circular}{A}.
That is, the red and blue lines in~\letterref{fig:experiments:circular}{A} show the average eigenvector centrality over 100 graphs drawn from each mixed CRGM, while the black lines show the eigenvector centralities expected by~\cref{claim:crgm-concentration}.

To illustrate the relationship between the differences in centrality and the difficulty of~\cref{prob:ranking}, we plot the empirical probability of correct pairwise ordering when $\gamma=0.8$ over $100$ trials when $m=1000$; see~\letterref{fig:experiments:circular}{B}.
For this experiment, we consider the same square-root filter excited with Gaussian white noise used in Section~\ref{sec:experiments:erdos-renyi}.
It is apparent that the probability of correctly ordering pairs of nodes is directly tied to their difference in centrality.
One can see a clear relationship between the probability of correct pairwise ranking plotted in~\letterref{fig:experiments:circular}{B} and the expected difficulty according to~\cref{thm:simple-nonasymptotic} plotted in~\letterref{fig:experiments:circular}{C}.

Comparing the elementwise perturbations of the leading eigenvectors when $\gamma \in \{0.2, 0.8\}$, we plot the mean minimum viable thresholds for both mixed CRGMs in~\letterref{fig:experiments:circular}{D}.
One can see that for all sample sizes $m$, the minimum viable threshold for $\gamma=0.2$ is smaller than that for $\gamma=0.8$ by a factor of approximately $1.5$.
That is, the elementwise perturbations of $\u$ when estimating $\hat{\u}$ were smaller for $\gamma=0.2$ than for $\gamma=0.8$.

However, plotting the completeness for both models reveals that for the same number of samples, the completeness of the partial ordering with the minimum viable threshold is higher when $\gamma=0.8$, as shown in~\letterref{fig:experiments:circular}{E}.
That is, although the mixed CRGM with $\gamma=0.2$ exhibits smaller elementwise perturbations as measured by the minimum viable threshold, the pairwise differences between centrality values for different nodes are also smaller, resulting in lower completeness than the ranking when $\gamma=0.8$.
Notice that this was expected since the more heterogeneous profile of centrality values when $\gamma = 0.8$ \cf{\letterref{fig:experiments:circular}{A}} should result in an easier ranking problem.

\begin{remark}[Localization affects perturbation bounds]\label{remark:flattening}
A more comprehensive characterization of this relationship between the absolute perturbation of the estimated eigenvector centrality $\uhat$ and the structure of the true centrality $\u$ is described in~\cite{roddenberry2020ranking}.
Indeed, it was demonstrated that, although the bound on the perturbation of $\uhat$ shrinks as $\u$ becomes increasingly delocalized, this bound does not shrink at the same rate as the range in values of $\u$ does, yielding increasingly harder ranking problems as $\u$ flattens.
\end{remark}

\subsection{Inferring centrality from U.S. Senate Republican voting records}\label{sec:experiments:senate}

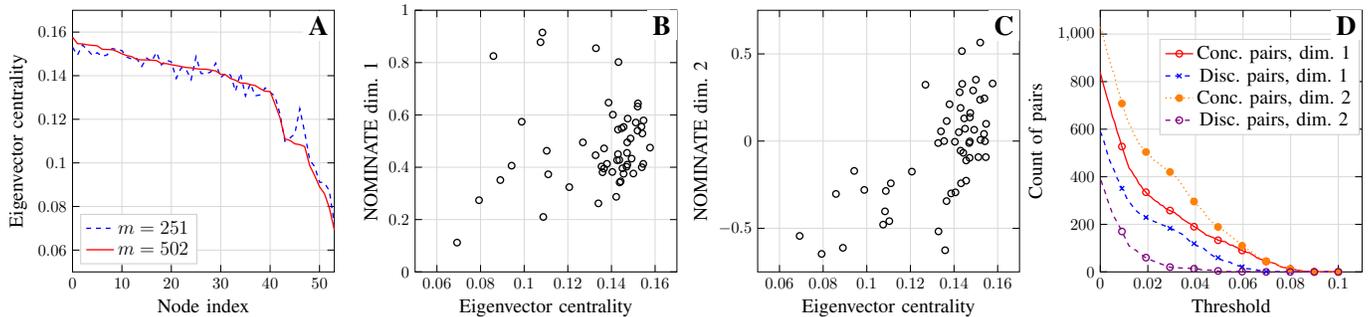
\begin{figure*}
  \centering
  \resizebox{\textwidth}{!}{\begin{tikzpicture}

  \begin{groupplot}[
    group style={
      group name=myplots,
      group size=4 by 1,
      horizontal sep=1.75cm},
    label style={font=\large},
    legend style={font=\large,draw=white!80.0!black},
    tick align=inside,
    tick pos=both,
    x grid style={gray!30},
    xmajorgrids,
    xminorgrids,
    xtick style={color=black},
    xticklabel style={
      /pgf/number format/fixed,
      /pgf/number format/precision=2,
    },
    y grid style={gray!30},
    ymajorgrids,
    yminorgrids,
    ytick style={color=black},
    yticklabel style={
      /pgf/number format/fixed,
      /pgf/number format/precision=2
    },
    scaled y ticks=false,
    width=0.4\textwidth,
    height=0.4\textwidth
    ]

    \nextgroupplot[
      legend pos={south west},
      xlabel={Node index},
      ylabel={Eigenvector centrality},
      no markers,
      xmin=0, xmax=53,
      ymin=0.05, ymax=0.17,
    ]
    \pgfplotsset{cycle list shift=-1}
    \addplot table[x=ranking, y=u_251] {./outdata/senate/ranking.csv};
    \addlegendentry{$m=251$};
    \addplot table[x=ranking, y=u_502] {./outdata/senate/ranking.csv};
    \addlegendentry{$m=502$};

    \nextgroupplot[
      xlabel={Eigenvector centrality},
      ylabel={NOMINATE dim. 1},
      xmin=0.05, xmax=0.17,
      ymin=0, ymax=1,
    ]
    \addplot+[only marks, black] table[x=centrality, y=nominate_1] {./outdata/senate/ranking.csv};
    
    \nextgroupplot[
      xlabel={Eigenvector centrality},
      ylabel={NOMINATE dim. 2},
      xmin=0.05, xmax=0.17,
      ymin=-0.75, ymax=0.75,
    ]
    \addplot+[only marks, black] table[x=centrality, y=nominate_2] {./outdata/senate/ranking.csv};

    \nextgroupplot[
      xlabel={Threshold},
      ylabel={Count of pairs},
      legend cell align=center,
      legend style={
        at={(0.99,0.89)},
        anchor=north east,
      },
      mark repeat=10, mark phase=1,
      ymin=0, ymax=1100,
      xmin=0, xmax=0.11,
    ]
    \addplot+[] table[x=tau, y=concordance1] {./outdata/senate/thresholds.csv};
    \addlegendentry{Conc. pairs, dim. 1}
    \addplot+[] table[x=tau, y=discordance1] {./outdata/senate/thresholds.csv};
    \addlegendentry{Disc. pairs, dim. 1}
    \addplot+[] table[x=tau, y=concordance2] {./outdata/senate/thresholds.csv};
    \addlegendentry{Conc. pairs, dim. 2}
    \addplot+[] table[x=tau, y=discordance2] {./outdata/senate/thresholds.csv};
    \addlegendentry{Disc. pairs, dim. 2}
    
  \end{groupplot}
  
  \node[below left,fill=white] at (myplots c1r1.north east) {\LARGE\textbf{A}};
  \node[below left,fill=white] at (myplots c2r1.north east) {\LARGE\textbf{B}};
  \node[below left,fill=white] at (myplots c3r1.north east) {\LARGE\textbf{C}};
  \node[below left,fill=white] at (myplots c4r1.north east) {\LARGE\textbf{D}};
\end{tikzpicture}

}
  \caption{
    Eigenvector centrality ranking estimation for Voteview data.
    (A) Estimated eigenvector centralities for $m=251$ and $m=502$.
    (B) Relationship between estimated eigenvector centrality and NOMINATE dimension 1.
    There is a weak monotonic relationship present ($\rho=0.221$).
    (C) Relationship between estimated eigenvector centrality and NOMINATE dimension 2.
    There is a strong monotonic relationship present ($\rho=0.623$).
    (D) Number of concordant and discordant pairs between $\uhat$ and each NOMINATE dimension for a range of thresholds $\tau$.
    For every $\tau$, the concordance between $\uhat$ and dimension 2 is greater than with dimension 1, and the discordance between $\uhat$ and dimension 2 is less than with dimension 1.
  }
  \label{fig:experiments:senate}
\end{figure*}

\begin{table}
  \centering
  \caption{Induced ranking from estimated eigenvector centrality for Republican senators.}
  \label{tab:experiments:senate}
  \input{figs/tikz/senate.table}
\end{table}

We consider roll-call voting records of Republican party members in the 114th U.S. Senate~\cite{voteview}.
The dataset provides $54$ individual senators' responses to $m=502$ roll-calls.
To interpret this as a graph signal, we consider each senator $i$ to be a node, and the signal value for roll-call $\ell$ to be
\begin{equation}\label{eq:senate-voting}
  y_{i}^{\Paren{\ell}}=
  \begin{cases}
    1 & i\text{ voted \emph{yea, paired yea, announced yea}}, \\
    -1 & i\text{ voted \emph{nea, paired nea, announced nea}}, \\
    0 & \text{otherwise}.
  \end{cases}
\end{equation}
Taking the covariance $\Cym$ of $\{\yl\}_{\ell=1}^m$, we estimate the eigenvector centrality via~\cref{alg:simple}.
The estimated centrality $\uhat$ has same-signed entries, as shown in~\letterref{fig:experiments:senate}{A}.
A partial description of the induced ranking of $\uhat$ is shown in~\cref{tab:experiments:senate}.
Considering the first $m=251$ roll-call votes, the estimated eigenvector centralities match that of the whole dataset ($m=502$) well, illustrated in~\letterref{fig:experiments:senate}{A}.
{That is, the general structure of the eigenvector centrality is apparently stable over subsets of the dataset.}

Since there is not an absolute known network among Republican senators, we compare the ranking induced by $\uhat$ for $m=502$ to the ideological positions of each senator as determined by the NOMINATE procedure~\cite{poole2005spatial}, which embeds each senator in $\Real^2$, indicating their ideological position in terms of economic policy (dimension 1) and other votes (dimension 2)~\cite{voteview}.
We then compare the ranking induced by $\uhat$ to that induced by each NOMINATE dimension using the Spearman correlation~\cite{cordernonparam}.

Quantitatively, the Spearman correlation between NOMINATE dimension 1 and $\uhat$ is $\rho=0.221\ (p=\num{0.109})$, while the Spearman correlation between NOMINATE dimension 2 and $\uhat$ is $\rho=0.623\ (p=\num{4.87e-7})$.
The marginal relationships between each of these dimensions and $\uhat$ is plotted in~\letterref{fig:experiments:senate}{B-C}.
The discovered correlation in rankings indicates that social conservatism (NOMINATE dimension 2) is a stronger indicator of centrality within the party than economic conservatism (NOMINATE dimension 1).

Additionally, we evaluate the behavior of~\cref{alg:threshold-ranking} for varying thresholds when comparing $\uhat$ and the NOMINATE dimensions.
For both dimensions, the minimum viable threshold leads to completeness of zero, due to highly non-monotonic relationships with $\uhat$ for some nodes.
Hence, we sweep over a range of thresholds, and count the number of concordant and discordant pairs in the partial order returned by~\cref{alg:threshold-ranking}, plotted in~\letterref{fig:experiments:senate}{D}.

As expected, in both cases the number of concordant and discordant pairs are monotonically decreasing with respect to $\tau$.
This is directly related to larger values of $\tau$ resulting in less complete partial orders.
Additionally, since dimension 2 correlates more strongly with $\uhat$, for every fixed threshold $\tau$, the number of concordant pairs between the partial order and $\uhat$ is greater for dimension 2 than dimension 1, and the number of discordant pairs is smaller for dimension 2 than dimension~1.

\section{Discussion}\label{sec:discussion}

In this work, we considered the blind centrality ranking problem, where we seek to rank the nodes in a graph by their eigenvector centrality without knowledge of the graph itself.
Instead, we observe a set of graph signals regularized by the graph structure via a graph filter.

Leveraging the shared eigenspaces of the covariance of these signals and the graph's adjacency matrix, we propose two simple algorithms for ordering the nodes based on their eigenvector centrality.
We show that these algorithms are correct in the asymptotic (infinite sampling) regime, and characterize the sampling requirements for correctness in terms of the true eigenvector centrality of the graph.
These characterizations are then demonstrated through extensive numerical experiments.

This work has many avenues for future research.
In the direction of inference of graph centralities, one could make stronger assumptions on the {spectral properties of the} graph filter (\eg Lipschitz smoothness) to obtain estimates of more involved spectral centrality measures, such as the Katz centrality.
Specifically, the Katz centrality yields a range of rankings parameterized by some scalar $\alpha$, interpolating the ranking induced by node degrees to the eigenvector centrality considered in this work.
Effectively handling the loss in phase information associated with the squaring of eigenvalues in the sample covariance matrix would be an exciting development in this direction.

In this and previous work on blind inference, spectral algorithms have been applied to the sample covariance function of the observed signals.
This is a valid choice under the assumed signal model \eqref{eq:system-model}, but breaks down in the presence of outliers or out-of-distribution behavior, perhaps due to multi-modal network processes.
An interesting problem would be the analysis of robust covariance estimation in this framework, either for the removal of noise or the separation of mixtures of graphs.
As an example, consider the centrality inference task for a temporal, in-person social network.
The network structure will typically be dynamic, switching between modes depending on whether people are, \eg at work or at home.
In such a situation, it would make more sense to consider the centrality structure of each distinct network mode, rather than the aggregation of the entire set of observations.

More broadly, the blind inference methodologies presented here and in past work provide a rich framework for the application of spectral graph theoretic algorithms to the analysis of a wide range of datasets, even where an explicit graph structure may not exist.
This is demonstrated in~\cref{sec:experiments:senate}, where a centrality ranking is observed in the voting patterns of U.S. Senators, although there is no explicit network connecting them and regulating their behavior.
Beyond centralities, one could conceivably apply methods such as spectral graph matching to the deanonymization of multiple datasets with shared covariates that lack known correspondence.
Moreover, we plan to extend the framework of blind inference of network features to higher-order relational structures represented by hypergraphs.

\appendices
\crefalias{section}{appendix}

\section{Proof of~\Cref{prop:sign-correction}}\label{app:sign-correction}

{Without loss of generality, assume that $\abrack{\u,\uhat}>0$ before sign-correction} \cf{step 5 in Algorithm~\ref{alg:simple}}.
We will show that if~\eqref{eq:sign-correction-inequality} holds, $\sgn{\abrack{\uhat,\bbone}}>0$, or equivalently, $0\leq\theta\paren{\uhat,\bbone\sqrt{n}}<\pi/2$.
{Then, if $\abrack{\u,\uhat}<0$ holds, contrary to this assumption, sign-correction will invert the estimate $\uhat\leftarrow -\uhat$ to positively align with $\u$.}

Since the vectors $\u,\uhat,\bbone/\sqrt{n}$ all have unit-norm, they lie on the surface of the sphere {${\mathbb S}^{n-1}$}.
Thus, the angles between these vectors obey the triangle inequality, since the angles between unit-norm vectors correspond to geodesics on the surface of the sphere.
That is,
\begin{equation}\label{eq:sign-correction-proof:triangle-ineq}
  \theta\Paren{\uhat,\bbone/\sqrt{n}}\leq\theta\Paren{\uhat,\u}+\theta\Paren{\u,\bbone/\sqrt{n}}<\pi/2,
\end{equation}
{where the second inequality would guarantee that $\sgn\paren{\abrack{\uhat,\bbone}}>0$.
Notice that the angle $\theta$ between two unit vectors $\x,\y$ can be written as $\theta\paren{\x,\y}=\arccos\paren{\abrack{\x,\y}}$.
Hence, for \eqref{eq:sign-correction-proof:triangle-ineq} to hold we require that}
\begin{equation}\label{eq:sign-correction-proof:triangle-ineq-alt}
  \Abrack{\u,\uhat}>\sqrt{1-\Abrack{\u,\bbone/\sqrt{n}}^2},
\end{equation}
from which~\eqref{eq:sign-correction-inequality} follows, as desired.

\small
\bibliographystyle{IEEEtran}
\bibliography{ref}

\end{document}